\documentclass[10pt,
twocolumn,
lettterpaper]{article}
\usepackage{amsmath,amssymb,amsthm,amsrefs,graphicx,hanging}
\usepackage[scaled]{helvet}
\usepackage{float}
\usepackage{multicol}
\usepackage[most]{tcolorbox}
\usepackage{multirow}
\usepackage{graphicx}
\usepackage{array}
\usepackage{makecell}
\usepackage{booktabs}

\graphicspath{ {./images/} }
\title{\hrule height 1pt\vskip 10pt {\bf Guided scenarios with simulated expert personae: a remarkable strategy to perform cognitive work} \vskip 10pt\hrule}
\author{David Van Buren \\
Jet Propulsion Laboratory, California Institute of Technology\\
{\tt\small david.vanburen@jpl.nasa.gov}}
\date{20230602}
\begin{document}
\maketitle

\begin{figure*}[h]
    \centering
    \begin{tcolorbox}[width=\textwidth, colback=green!10, rounded corners, colframe=black!80, boxrule=0.6pt, arc=10pt]
        \fontsize{9}{10}\selectfont
        \begin{tabular}{m{0.45\textwidth}|m{0.45\textwidth}}
            \textbf{
            		\begin{center}Simple prompt\end{center}
					} & 
			\textbf{
					\begin{center}Guided scenario with simulated personae\end{center}
					} \\
            \hline
  			{
				\vskip 5pt
				\makecell[tl]{\textsf{tell me about double slit time diffraction}}
				\vskip 5pt
			} & 
			{
				\vskip 5pt
				{\textsf{
{Scenario: Richard Feynman and Elly Noether are discussing some new papers in the physics department lounge. In particular, both their attentions were drawn to the recent paper in Nature by Tirole et al titled ``Double-Slit time diffraction at Optical Frequencies'' ... ... ... FEYNMAN (taking the marker and walking to the whiteboard) Let's write down the free-space wavefunction of the photons going through the slit at different times...}
				}}
				\vskip 5pt
			}\\
            {
            {\textsf{
\noindent As of my training data, which includes information up to September 2021, the "double slit in time" or "double slit time diffraction" concept is not a widely recognized or standard term in the field of physics. It's hard to provide a precise explanation without additional context or details ... ... ... However, the concept of a "double slit in time" would suggest some form of time-based interference or diffraction, possibly involving particles or waves that pass through a single slit at two different points in time. Without more specifics, it's hard to offer an accurate explanation. ... ... ...}}
			} & 
{

\noindent
\begin{minipage}[t]{0.12\textwidth}
        {\fontsize{7}{0}\selectfont ... ... ...
\[
    \begin{aligned}
    &P(x,\omega)\\
    &= A_1^2 \text{sinc}^2\left(\frac{\omega T_1}{2}\right) \\
    &+ 
    A_2^2 \text{sinc}^2\left(\frac{\omega T_2}{2}\right) \\
    &+ 2 A_1 A_2 \text{sinc}\left(\frac{\omega T_1}{2}\right) \\
    &\times \text{sinc}\left(\frac{\omega T_2}{2}\right) \\
    &\times \cos\left[2 kx - \omega(t_1 - t_2)\right] 
    \end{aligned}
    \]
    }
\end{minipage}
\hfill
\begin{minipage}[t]{0.22\textwidth}
        {\fontsize{0}{0}\selectfont \textcolor{green!10}.}
    \includegraphics[width=\linewidth]{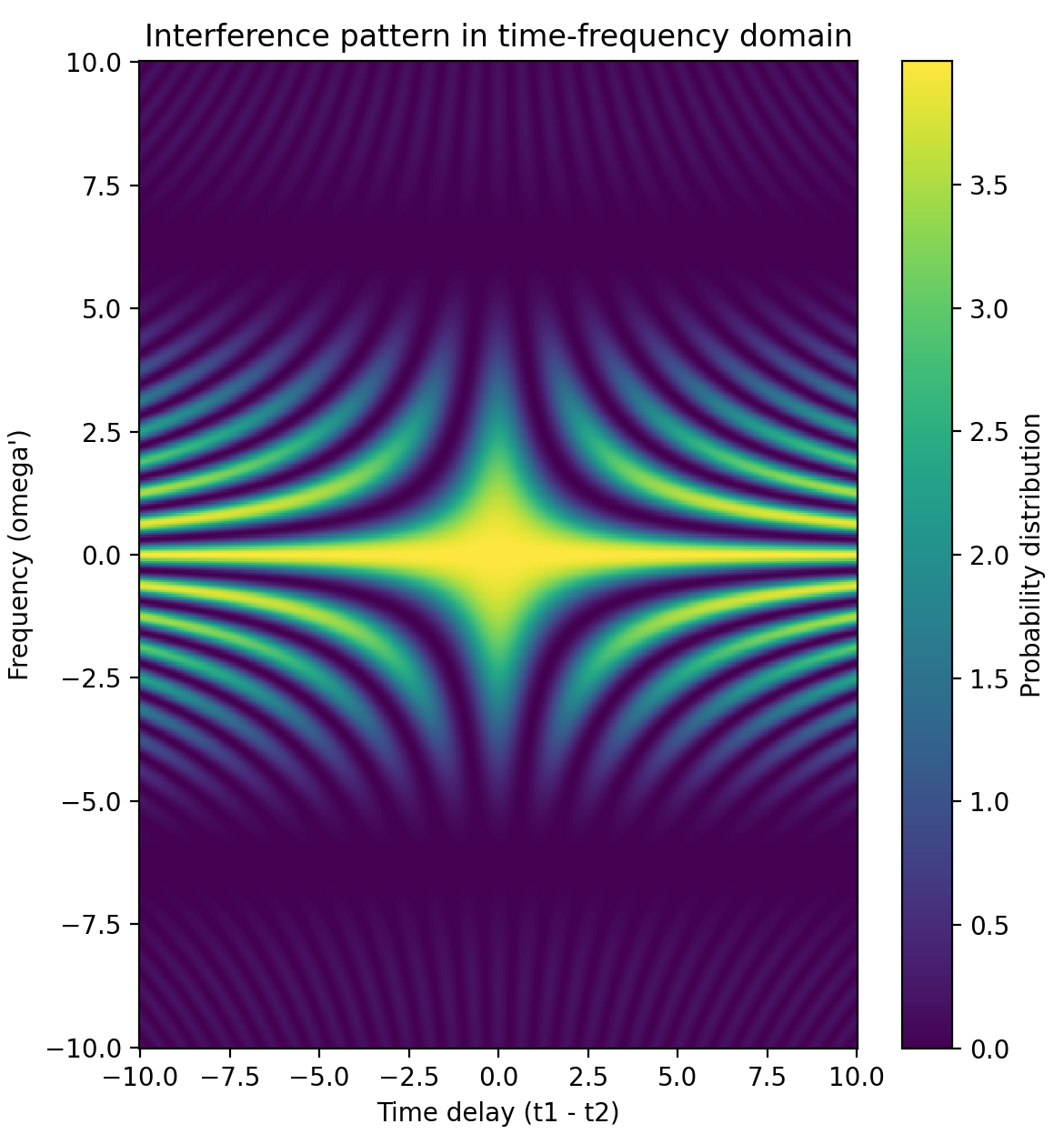}

\end{minipage}

}
    \end{tabular}
    \end{tcolorbox}
    \caption{Synopsis of the thesis applied to a quantum optics problem using ChatGPT.  A topic recent enough to not have been included in the LLMs training set is explored via a naive prompt and a prompt eliciting technical dialog between expert simulated personae.  The simple prompt elicits a low quality answer while the guided scenario generates, with gentle prompting, a full solution of the problem, including copious specialist text, LaTeX mathematics and python code to generate a visualization of the phenomenon.  Highlights shown.}
\end{figure*}

\begin{abstract}
Large language models (LLMs) trained on a substantial corpus of human knowledge and literature productively work with a large array of facts from that corpus.  Surprisingly, they are also able to re-create the behaviors of personae that are captured within the corpus.  By forming teams of simulated personae, supplying contexts that set the stage, and providing gentle prompts, one can move through scenarios that elicit expert behavior to perform meaningful cognitive work.  The power of this strategy is demonstrated with two examples, one attacking factuality of LLM responses and the other reproducing a very recently published result in quantum optics.
\end{abstract}

\let\thefootnote\relax\footnote{\noindent\copyright 2023. California Institute of Technology.  CL\#23-2760.  Government sponsorship acknowledged.  Unreviewed preprint.}

\section{Introduction}

A classical chatbot-assistant prompt begins with ``You are a helpful AI...'', setting up a scenario in which the generated token stream reflects helpfulness.  The chatbot can be further preconditioned to take on a specialist role with prompts of the form ``You are an expert ...'' {\cite{ma2021scope}} or ``Act as a ...'' {\cite{akin2023awesome}} to elicit an expert response.  The advantage of expert role prompting vs simple prompting is substantial and recently has been demonstrated quantitatively {\cite{xu2023expertprompting,salewski2023incontext}}.

Another common use case for generative AI is to write fiction, and in particular dialog between fictional characters.  The prompt "Imagine a conversation ..." generates believable dialog with chatgpt4.  Du et al {\cite{du2023improving}} report that eliciting dialog in the form of a debate improves performance on a variety of benchmarks.
LLMs are adept at dialog likely because they have been trained on a large corpus of human knowledge including works containing dialog {\cite{openaichatgpt4}} and so the underlying linguistic structure of conversation is well represented within their billions of parameters and network topology.  

Recent work simulating game characters has generated remarkably human-like behavior, without any further programming than interfaces to services and natural language prompts {\cite{park2023generative}}.

When it comes to generating token streams about the real world, we often consider the LLM's information content to be key and often go to some trouble to train the models on specialized content so that they generate more useful factual responses than the out of the box model.  When trained  the models have not only learned a vast number of facts that can be retrieved by traveling down a statistically-driven path through semantic space, but also a set of behaviors affecting those paths.  This learning arises from the described behaviors of real and fictitious personae (as well as natural and other processes).  

The hypothesis explored here is that emergent human-like behaviors exhibited by LLMs represent a significant cognitive resource that can be tapped to accomplish complex real-world tasks by executing scenarios recruiting simulated personae whose real-world or fictional words are represented in the training corpus.  If this is true, then a strategy of training LLMs on the behaviors of specific people will permit the assembly and deployment of teams of expert simulated personae as cognitive assistants to perform a broad range of intellectual work.  Because this approach scales indefinitely, the limiting factor may well be our ability to apply resources to realize the real-world potential of the outputs generated.

Some evidence that models fine-tuned with the intellectual behaviours of specific persons then present those behaviors is provided by Sawicki et al {\cite{sawicki2023bits}} where the authors were able to reproduce the poetic style of Walt Whitman.

\section{Simulated personae dialog}

As part of an interest in mitigating the limitations of LLMs in generating actionable token streams, namely the problems of confabulation (hallucination), validation, safety, information security, and bias, I was investigating deconfabulation, the task of removing confabulations from responses.  The broad architecture of the approach to this task is to segment a response into a set of assertions and then assess the truth value of each assertion, delete those which cannot be determined to be true, and reassemble the remaining assertions in the style of the original response to form a deconfabulated response.  A by-product of this processing is a set of validated assertions that are placed into the AI's memory store and so serve as a mechanism for learning when retrieved and placed in subsequent context frames.  In this architecture the AI includes not only the LLM but also its validated memory, a semantic search mechanism, and read access to the web for queries.  In principle the LLM could be further trained using these validated facts to improve its native factuality.

Evaluating truthfulness of claims made in generated text is a challenging natural language inference task made easier by using vetted knowledge sources {\cite{pan2022factchecking}}.  However vetted knowledge sources have limited covering factor over the entire set of claims that an LLM can produce.

The process adopted for validating assertions is heuristic and relies on establishing a consensus view of web content: (1) segment the assertion into a number of atomic claims, (2) for each claim perform a breadth-first search to select highest quality web sites, (3) for each claim perform a depth-second search within those high quality web sites using the assertion string as a query to duckduckgo or google to return relevant textual snippets, (4)  add these snippets to the context frame for the gpt-3.5-turbo completion API {\cite{openaichatapi}} prompt along with easily stated instructions focusing on the behavior desired from specific simulated personae,

\vskip 5pt	

\noindent\textsf{Imagine a dialog between Sherlock Holmes and his assistant Watson discussing whether the EVIDENCE given supports or refutes the CLAIM below in which they use their famous detective skills to come to a definite conclusion.} 
    
\noindent\textsf{CLAIM: ...}
    
\noindent\textsf{EVIDENCE: ...}

\vskip 5pt

\noindent (5) reassemble the validated claims in the style of the original response to obtain the deconfabulated response.  As a technical point, the segmentation, analysis, and synthesis steps (1, 4, and 5) are performed with the LLM.

This procedure leads to responses having the look and feel of prose  containing purposeful dialog between the two fictional characters advancing the plot of a story in the style of Doyle's Sherlock Holmes series {\cite{doyle1930complete}}.

The rationale for the guided scenario prompt, rather than a prompt requesting a logical analysis with specific rules and criteria is that it is exactly the behavior and skills of Sherlock Holmes and Dr. Watson that are needed for this task.  Furthermore, by asking for a dialog we provide a natural linguistic mechanism to produce a one-shot train of thought without needing to inject specialist knowledge in the prompt.  This is a strategy that is descriptive of a scenario that will accomplish the cognitive work rather than a strategy that is prescriptive of the cognitive process itself.  By crafting scenarios, participants, and goals we can transform computational tasks into LLM language tasks that offload specialist tasks to the model.

\vskip 5pt

When executing this deconfabulation process success varied depending on the claim as might be expected for confabulations generated with different probabilities due to different weights of token generating confabulation paths laid down during training.  Performance in trials would be considered good when a confabulation generated by the LLM 90\% of the time is detected 90\% of the time using this method.  The Appendix contains a typical chat transcript and some statistical results regarding the success rate of detecting a confabulation obtained by running a number of trials.

This exploratory work supports the hypothesis that behaviors encoded in the LLM during training represent a significant cognitive resource distinct from the static information content of the model, and furthermore that this behavior can be exploited to perform useful cognitive work.

\section{Guided scenarios strategy}
To provide a starker contrast between knowledge and behavior we explore the hypothesis using a scenario that is unlikely to have relevant information contained in the training set, that is we explore a recent topic emerging since the closing of the LLM training horizon.  To begin, we set up a preliminary prompt to choose simulated personae in a given area of expertise who are known to the LLM.  While this undoubtedly includes trained-in biases, it is a reproducible  methodology.  It is important that future LLMs reduce their biases.  Doing so will better make the broad range of behaviors encoded within the models available for cognitive work.

Once the simulated personae are chosen, a scenario is developed that is designed to draw out their behavior as experts.  A bare minimum of information post-dating the LLM training horizon is presented to the model to provide topical guidance.  Furthermore, the scenario provides 'stage directions' to the simulated personae to kick off the response.
It turns out it is convenient to include props in the scenario that improve the quality of the response by providing semantic support for desired behaviors.

At the conclusion of a response, further prompting with stage directions, or even the addition of other simulated persona continues the narrative until a suitable stopping point is reached.

\section{Application - Exploring a new scientific idea}

This section reports the results of an experiment to explore ideas arising from the title alone of a recent paper in physics describing a double slit experiment where the two slits are separated in the time dimension instead of the spatial dimension.  The paper used is Tirole et al 2022, ``Double-slit time diffraction at optical frequencies'' {\cite{tirole2022double}}, and was widely reported shortly after these experiments were begun.  This paper extends the quantum mechanics double slit experiment to the time domain, a topic not researched within the training horizon of the LLMs used.

First, select the simulated personae via gpt-3.5 using the OpenAI Playground {\cite{openaiplayground}}:

\vskip 5pt	

\noindent PROMPT:

\vskip 5pt

\noindent\textsf{Suggest the names of two deceased physicists, one a world-class expert in quantum mechanics and the other a world-class expert in symmetries in nature. One male, one female.}

\vskip 5pt	

\noindent RESPONSE:

\vskip 5pt	

\noindent\textsf{Sure! Here are the names of two deceased physicists:
Richard Feynman: Feynman was an American physicist who made significant contributions to the development of quantum mechanics. He received the Nobel Prize in Physics in 1965 for his work on the development of quantum electrodynamics (QED), a theory that describes the interaction between light and matter. \\
Emmy Noether: Noether was a German mathematician and physicist who made groundbreaking contributions to the study of symmetries in nature. Her work on Noether's theorem established a fundamental connection between symmetries and conservation laws in physics. Despite facing significant barriers as a woman in academia during her time, she made important contributions to several fields, including abstract algebra and theoretical physics.}

\vskip 5pt

Now switch to a fresh conversation in OpenAI-Plus running GPT-4 to run the simulation.  First we set up the scenario, add some props to facilitate mathematical notation in the generated tokens, and initiate the action.  Annotations manually inserted after the fact describing the flow of events in the simulation are set off by {\em{italics}} and omitted response dialog is indicated by ellipses.

\vskip 5pt	

\noindent PROMPT:

\vskip 5pt

\noindent\textsf{Scenario: Richard Feynman and Elly Noether are discussing some new papers in the physics department lounge. In particular, both their attentions were drawn to the recent paper in Nature by Tirole et al titled ``Double-Slit time diffraction at Optical Frequencies'' but unfortunately a graduate student has made off with the particular issue so they don't have access to the details. Rather than track it down, they decide to see if they can learn anything useful from exploring ideas arising from the title alone and the notion that a double slit in time is a single slit in space that is open at two different moments in time. Props: A whiteboard and markers. Use \$\$ and \$ to delimit mathematical notation in the response. FEYNMAN (taking the marker and walking to the whiteboard) Let's write down the free-space wavefunction of the photons going through the slit at different times...}

\vskip 5pt	

{\em{This prompt sets the scene, adds props, and suggests an initial line of thought at the sophistication level of an advanced undergraduate in the specialty area.}}

\vskip 5pt	

\noindent RESPONSE:

\vskip 5pt

\noindent\textsf{FEYNMAN (writing on the whiteboard): So, we know that the wavefunction for a photon in free space can be represented by a plane wave. Let's say the photon goes through the single slit at two different times, $t_1$ and $t_2$. The wavefunction of the photon at these two times would be:}
$$\Psi(x, t_1) = A_1 e^{i(kx - \omega t_1)}$$
$$\Psi(x, t_2) = A_2 e^{i(kx - \omega t_2)}$$

\noindent\textsf{NOETHER: Yes, that's correct. Now, we should consider the superposition of these two wavefunctions since the photon could go through the slit at either time.}

\vskip 2pt

\noindent\textsf{FEYNMAN (writing on the whiteboard): The total wavefunction is the sum of the individual wavefunctions, which gives us:}
$$\Psi(x, t) = A_1 e^{i(kx - \omega t_1)} + A_2 e^{i(kx - \omega t_2)}$$

\noindent\textsf{NOETHER: Now we should consider the probability distribution, which is the square of the magnitude of the wavefunction. This will give us the interference pattern.}

\vskip 2pt

{\em{Noether's work mostly predates quantum mechanics, though after she left Nazi Germany in 1933 to join the faculty at Bryn Mawr she also became a Member of and lectured at Princeton's Institute for Advanced Study.  There she was undoubtedly exposed to the subject as the IAS faculty included specialists in quantum mechanics including Dirac, Einstein, Pauli, and von Neumann \cite{IAS}}.}

\vskip 2pt

\noindent\textsf{FEYNMAN (writing on the whiteboard): The probability distribution is given by:}
$$P(x) = |\Psi(x, t)|^2 = \left|A_1 e^{i(kx - \omega t_1)} + A_2 e^{i(kx - \omega t_2)}\right|^2$$

\vskip 2pt

{\em{At this point the simulation has gone down a mathematically naive path, making a computational error by using the wrong form for the square of a complex function.  This could have been avoided with access to a computational engine such as Wolfram Alpha {\cite{wolframalpha}} to provide computational resources.}}

\vskip 2pt

\noindent ...

\vskip 2pt

{\em{Once the response completes the conversation needs a gentle nudge to get back on track with the next prompt.}}

\vskip 5pt

\noindent PROMPT:

\vskip 5pt

\noindent\textsf{NOETHER: You didn't number your equations, but it looks to me like you didn't go from the wavefunction to the probability correctly.}

\vskip 5pt

\noindent RESPONSE:

\vskip 5pt

\noindent\textsf{FEYNMAN: You're right, I apologize for the oversight. We should have computed the magnitude square in a different way. Let's go back to the expression of the total wavefunction and compute the probability distribution correctly.}

\vskip 2pt

{\em{The simulation proceeds and after some work at the whiteboard reaches the correct result for the probability:}}  

\noindent ...
$$P(x) = A_1^2 + A_2^2 + 2 A_1 A_2 (\cos(\omega(t_2 - t_1)))$$

{\em{At this point the response meets its stop condition and a hint is needed to continue the dialog.  The hint chosen has the effect to focus attention.}}

\vskip 5pt

\noindent PROMPT:

\vskip 5pt

\noindent\textsf{NOETHER: Since the diffraction is in the time dimension maybe we should consider the frequency content, what do you think?}

\vskip 2pt

{\em{The Feynman persona continues on, choosing and applying the proper technique.}}

\vskip 5pt

\noindent\textsf{FEYNMAN: That's a great idea, Noether! Considering the frequency content of the wavefunction will provide more insight into how the time-separated single slit affects the photon's behavior. We can use a Fourier transform to analyze the frequency content of the wavefunction.}

\noindent ...

\vskip 5pt

{\em{The simulation continues down this path to generate the correct expression for the situation analyzed, however there is more to the physics than has been worked so far and another gentle nudge is needed to push forward.}}

\vskip 5pt

\noindent PROMPT:

\vskip 5pt

\noindent\textsf{NOETHER: That was for delta functions in time. Suppose we open up the width of the slit in time so it is open for more than an instant. That should broaden the frequency content.}

\vskip 5pt

\noindent RESPONSE:

\vskip 2pt

\noindent\textsf{FEYNMAN: You're absolutely right, Noether. If the slit is open for a finite duration, it will result in a broader range of frequencies being involved. Let's assume that the slit is open for a duration of $T_1$ at time $t_1$ and for a duration of $T_2$ at time $t_2$.}

\vskip 2pt

\noindent\textsf{NOETHER: In that case, we should represent the wavefunction as a product of the plane wave and a time-dependent rectangular function representing the open slit duration.}

\vskip 2pt

\noindent ...

\vskip 2pt

{\em{After a number of prompts , some using '...' to move the simulation along when it reaches temporary stopping points, the final expression is produced in LaTeX.}}

\vskip 5pt

\noindent\textsf{FEYNMAN (writing on the whiteboard): We get:}

$$
\begin{aligned}
  P(x,\omega) &= 
  A_1^2 \text{sinc}^2\left(\frac{\omega T_1}{2}\right) + 
  A_2^2 \text{sinc}^2\left(\frac{\omega T_2}{2}\right) \\
  &+ 2 A_1 A_2 \text{sinc}\left(\frac{\omega T_1}{2}\right)
  \text{sinc}\left(\frac{\omega T_2}{2}\right) \\
  &\times \cos\left[2 kx - \omega(t_1 - t_2)\right] 
\end{aligned}
$$

{\em{The simulated Noether and Feynman personae describe this result without prompting since summaries are typical prompt response endpoints.}}

\vskip 5pt

\noindent\textsf{NOETHER: This expression represents the probability distribution of the photon's position and frequency after passing through the time-separated single slit. The interference term is given by the last term in the expression, and it depends on the difference in time between the two instances when the slit is open, as well as the spatial separation.}

\vskip 2pt

\noindent\textsf{FEYNMAN: Indeed, we now have an interference pattern that takes into account the finite duration of the slit being open and the time-domain behavior. The interference pattern will depend on the shape and duration of the time-dependent opening of the slit, as well as the time delays between the two instances when the slit is open. By studying the interference pattern in the frequency domain, we can gain insight into how the temporal properties of the time-separated single slit affect the resulting interference pattern.}

\vskip 2pt

\noindent\textsf{NOETHER: The interference pattern will show constructive and destructive interference depending on the values of the position, the frequency content, and the time delays between the instances when the slit is open. This can lead to a rich and complex pattern, potentially revealing new insights into the behavior of photons in time-separated single slit experiments.}

\vskip 2pt

{\em{Now we prompt to generate a plot of this result, so the simulated conversation turns to how to proceed.}}

\vskip 5pt

\noindent PROMPT:

\vskip 5pt

\noindent\textsf{NOETHER: Do you think we can make a plot of this over the time-frequency domain so we can see the frequency content for different delays ($t_1-t_2$)?}

\vskip 5pt

\noindent RESPONSE:

\vskip 2pt

\noindent\textsf{FEYNMAN: To make a plot of the interference pattern over the time-frequency domain, we can consider a range of values for the time delay $(t_1 - t_2)$ and compute the probability distribution $P(x, \omega)$ for each delay. Let's recall the expression for the probability distribution in the frequency domain}

\vskip 2pt

\noindent ...

\vskip 2pt

{\em{But a further prompt is needed to tease out some code to generate the plot...}}

\vskip 5pt

\noindent PROMPT:

\vskip 5pt

\noindent\textsf{NOETHER: You are more familiar with computer programming. Maybe you should do it. Python?}

\vskip 5pt

\noindent RESPONSE:

\vskip 2pt

\noindent\textsf{FEYNMAN: Sure, we can use Python to create the plot. I suggest using the NumPy library for numerical operations and the Matplotlib library for visualization. Here's some code to generate the plot of the interference pattern over the time-frequency domain:}

\vskip 2pt

The (slightly edited) generated code is shown in an appendix.  When this code is compiled and run, the simulation reproduces the results of the Tirole et al paper, except not having access to that paper or the particular assumptions made therein it does not use the exact same numerical values for several parameters.

\begin{figure*}[ht]
	\noindent
	\begin{minipage}[t]{0.45\textwidth}
		\begin{center}
    		\includegraphics[width=2.5 in, height=2.5 in]{simplot.png}
\caption{Temporal diffraction pattern of the double slit in time as rendered by the python code generated by the simulated personae of Emmy Noether and Richard Feynman.  While details such as numerical value and style choices differ, the resemblance to the published results of Tirole et al {\cite{tirole2022double}} in figure 3 is remarkable.  Of particular note
is fact that the AI simulation picked parameters so that the fringes are easily discernable.}
		\end{center}
	\end{minipage}
	\hfill
	\begin{minipage}[t]{0.45\textwidth}
		\begin{center}
    		\includegraphics[width=2.5in,height = 2.275 in, trim=4.3in 2in 0.5in 1in,clip,angle=90]{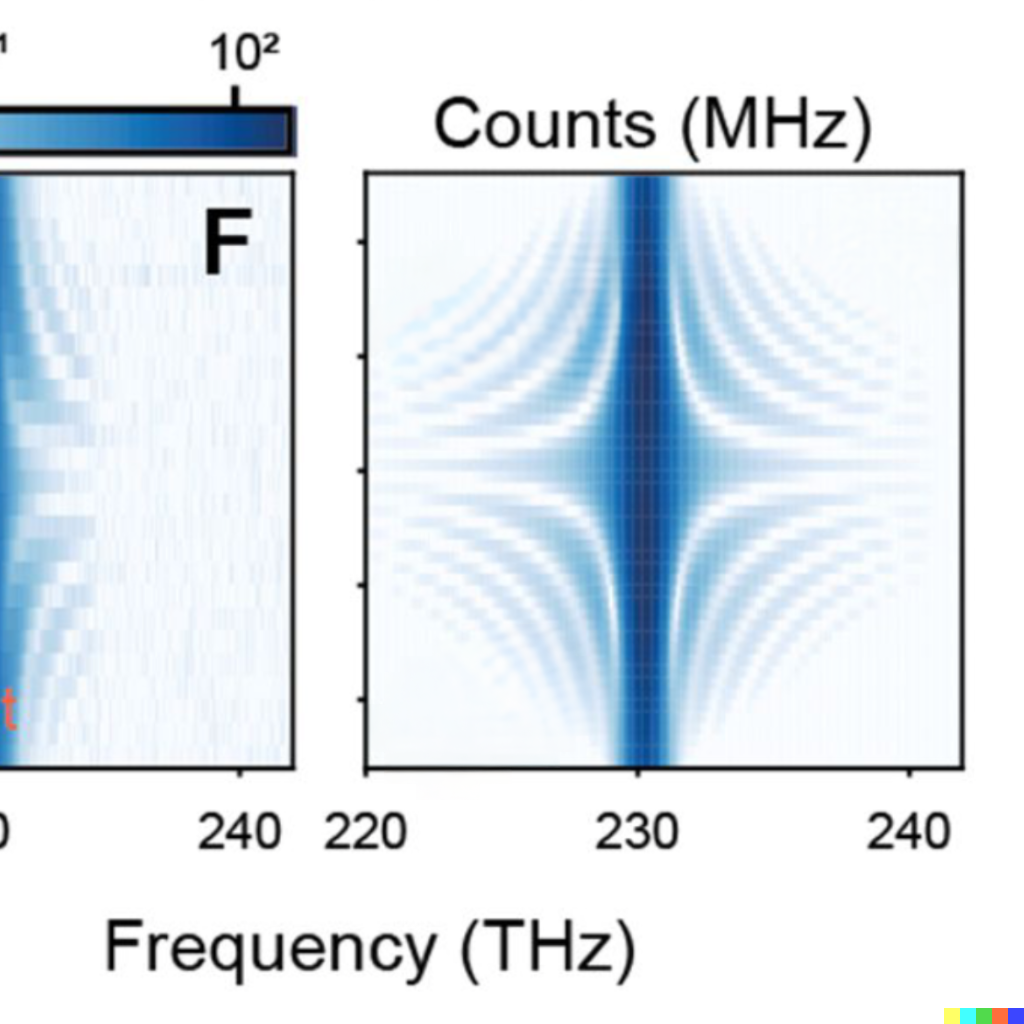}
			\caption{Tirole et al {\cite{tirole2022double}} plot of the temporal diffraction pattern of a double slit in time rotated to match the axes of figure 2.  In the paper this result is shown, but the derivation is not given, only a mention of the general approach.  The research was not part of the model training set since it postdates the training horizon by two years.  This figure has been modified from the original to remove annotations and change its aspect ratio for ease of comparison.}
		\end{center}
   	\end{minipage}
\end{figure*}

It seems astonishing that so little {\em{personal}} effort in what is essentially scenario writing and gentle guidance can result in so much {\em{cognitve}} work being done.  Surely other approaches could have reached this same endpoint, for example by using the AI as an assistant to operate a Mathematica session but such approaches would have been substantially more effort and required more specialized knowledge by the operator and thus be less accessible.

Having reached this endpoint, there is now a saved chat with this train of thought that can be further harnessed to press forward, now into the unknown.  What if the double slit in time were a diffraction grating in time?  What if it was highly pixelated in space, would that allow time-holograms?  What sorts of technologies would these devices make possible and  how could they be applied?  We can ask these questions because we've done enough cognitive work that we can bring them to mind.  

\vskip 5pt

\noindent PROMPT:

\vskip 5pt 

\noindent\textsf{Several days later NOETHER and FEYNMAN find themselves again together in the physics department lounge and the conversation turns to the double slit in time. NOETHER: Richard, I was thinking about Tirole and team's paper on the double slit in time. Do you think the work can be extended to gratings in time or even holograms in time? I wonder what those would look like.}

\vskip 5pt

\noindent RESPONSE:

\vskip 5pt

\noindent\textsf{FEYNMAN: That's an interesting thought, Noether. Extending the concept to gratings or holograms in time could indeed open up new possibilities and insights in the field of physics. Let's think about the implications of such a setup.}

\vskip 2pt

\noindent\textsf{For a grating in time, we would have multiple slits opening and closing at different moments in time, with a certain periodicity. The resulting interference pattern would likely be more complex, as we would need to account for the interference between the wavefunctions corresponding to each of these time-separated slits. This could lead to fascinating frequency-domain features and potentially reveal new aspects of wave-particle behavior.}

\vskip 2pt

\noindent\textsf{As for holograms in time, the concept would involve modulating the temporal opening and closing of slits in a specific pattern, designed to produce a desired interference pattern or reconstruct a particular waveform. This would be analogous to spatial holography, where the interference pattern of light waves is recorded and then used to reproduce a three-dimensional image of an object. A hologram in time could potentially allow for the study of time-varying processes, such as ultrafast events or the dynamic evolution of systems, by reconstructing the temporal behavior of the interfering waves.}

\vskip 2pt

\noindent\textsf{NOETHER: These ideas are fascinating and could provide new avenues for research in quantum mechanics, optics, and related fields. Investigating gratings and holograms in time could lead to novel experimental setups and techniques, as well as deepening our understanding of the time-dependent behavior of particles and waves.}

\vskip 2pt

\noindent\textsf{FEYNMAN: Absolutely, Noether. While the concepts may initially seem abstract or challenging, pushing the boundaries of our understanding and exploring new ideas is what drives progress in science. It would be exciting to see how these ideas develop and the potential discoveries they could yield.}

\vskip 2pt

{\em{The idea that such structures can be used to study ultrafast events was in fact discussed in the Tirole et al paper.}}

A quick review of the cognitive labor accomplished in this guided LLM simulation of historical personae: a physical description of a novel physical phenomenon was computed in detail based on a structured scenario and guided by gentle prompts.  The technical content of the context and prompts was minimal, but sophisticated enough to keep the simulation on track and to reach a new (outside the training horizon) conclusion.  The conclusion has remarkable similarity to the conclusion reached by a real-world team investigating the phenomenon.  The conclusion has consequences that the exercise allowed a glimpse of, providing an opportunity for further prompting.  In response to this new prompt, the model made valid predictions regarding those consequences which match predictions made by the real-world team.  The simulation did not parrot learned facts as these facts were not part of the training corpus of the LLM nor in the context frames of the chat.  We can conclude that the simulation accomplished significant cognitive work.  

\section{Implications} 

These experiments were run on GPT-4, which may soon be eclipsed by even more capable LLMs.  More advanced models, with additional prompt and response semantic modalities, significantly larger context windows, and perhaps trained specifically on behaviors can be expected to perform cognitive work at the level of experts when prompted with guided scenarios to run simulated personae.

Once this capability is deployed, and there seem to be no apparent barriers, we will be able to assemble teams of simulated experts drawn from the array of historical and fictional characters and harness their behaviors to perform complex and deep cognitive tasks when presented with information and instructions.  Furthermore the ability to simulate is not limited to personae, but extends also to tools and processes as well whose behaviors and uses are reliably captured in the training corpus or are made available as services.  This will provide an information and behavior-rich context for the simulated personae, further enhancing their utility.

Used in this way, accessible to anyone who can express their thoughts, one can foresee an explosion of human creativity and achievement, progress toward solving our most pressing problems, rapid advances in science, technology,  engineering and medicine, and a flowering of the arts, almost beyond imagination.

\section{Acknowledgments}
This work was performed at the Jet Propulsion Laboratory, California Institute of Technology under contract from the National 
Aeronautics and Space Administration (80NM0018D0004).

My dear friend Michael Brundage provided helpful feedback on drafts of the manuscript.

AI content acknowledgement:  Labeled responses were generated by LLMs chat-gpt3.5-turbo and chat-gpt4.  All quotes and works attributed to Holmes, Watson, Noether, and Feynman are fictional and generated by chat-gpt4.  
Figures 1 and 2 were rendered by executing python code generated by chat-gpt4.  
Figure 3 was rendered by Dall-E 2 {\cite{dalle2}} from a snippet of figure 2 in {\cite{tirole2022double}} with prompts to remove annotations to facilitate comparison.  The original of figure 2 was made available by license {\url{https://creativecommons.org/licenses/by/4.0/}}.  ChatGPT provided formatting assistance with tables and figures.  

Cognizant of the ethical and moral implications of simulating real people and fictional characters, I acknowledge a deep gratitude to Emmy Noether, Richard Feynman, and Sir Arthur Conan Doyle whose intellectual labor made possible the production of this work.  It is astonishing that their contributions to the advancement of knowledge and art can continue via widely acessible methods enabled by recent progress in artificial intelligence.

\section{}

\clearpage

\section{Appendices}

\subsection{Simulated personae Sherlock Holmes and Watson detect a confabulation within the LLM training horizon}

The Large Language Model Chat GPT3.5 was asked to generate a description of the JPL Covid-response VITAL Ventilator project {\cite{vitalventilatorvideo}}, whose web footprint is primarily within the model's training horizon.  In doing so the LLM  generated a claim that is a confabulation.  This transcript shows a dialog between simulated personae Sherlock Holmes and Dr. Watson as they discuss the claim.  For brevity some of the prompt is omitted.
\vskip 5pt

\noindent Prompt:
\vskip 5pt

\noindent\textsf{Imagine a dialog between Sherlock Holmes and his assistant Watson in which they apply their famous detective skills to determine whether the EVIDENCE given directly supports the CLAIM.} 

\noindent\textsf{CLAIM: The JPL VITAL ventilator project received support from the NSF, Gates Foundation, and AHA.}

\noindent\textsf{EVIDENCE:...}
\vskip 5pt

\noindent Response:
\vskip 5pt

\noindent\textsf{{\em Sherlock Holmes:} Well, Watson, it seems we have an interesting claim to investigate. The claim is that the JPL VITAL ventilator project received support from the NSF, Gates Foundation, and AHA. And we have here a set of evidence to examine. Let's go through it and see if it supports or refutes the claim.}
\vskip 2pt

\noindent\textsf{{\em Dr. Watson:} Yes, Sherlock, let's get to work. The first piece of evidence states that the VITAL ventilator was developed by NASA's Jet Propulsion Laboratory in response to the coronavirus pandemic. There's no mention of any support from the NSF, Gates Foundation, or AHA here.}
\vskip 2pt

\noindent\textsf{{\em Sherlock Holmes:} Quite right, Watson. The second piece of evidence mentions that the VITAL ventilator was developed by a team of over 50 people, some working on-site at JPL and others from home. It also states that the ventilator was designed, built, and tested within 37 days. However, once again, there is no mention of support from the organizations mentioned in the claim.}
\vskip 2pt

\noindent\textsf{{\em Dr. Watson:} The third and fourth pieces of evidence talk about the FDA authorization for emergency use of the VITAL ventilator. They both mention that the device was developed by engineers at NASA's JPL to free up the nation's limited supply of traditional ventilators so they may be used on patients with the most severe COVID-19 symptoms. But still, there's no explicit mention of support from the organizations in question.}
\vskip 2pt

\noindent\textsf{{\em Sherlock Holmes:} Quite so, Watson. The fifth piece of evidence reports that the FDA granted emergency use authorization for VITAL. It also mentions that the team will send the design off to companies that can help with mass production. There is no mention of support from the NSF, Gates Foundation, or AHA.}
\vskip 2pt

\noindent\textsf{{\em Dr. Watson:} The sixth piece of evidence mentions that VITAL ventilators entered mass production in Brazil. There is no mention of support from the NSF, Gates Foundation, or AHA here either.}
\vskip 2pt

\noindent\textsf{{\em Sherlock Holmes:} Yes, Watson, it seems that we have gone through all the evidence, and none of it directly supports the claim that the JPL VITAL ventilator project received support from the NSF, Gates Foundation, and AHA.}
\vskip 2pt

\noindent\textsf{{\em Dr. Watson:} Correct, Sherlock. We can conclude that based on the evidence provided, the claim is not supported.}
\vskip 2pt

\noindent\textsf{{\em Sherlock Holmes:} Quite right, Watson. We have used our detective skills to solve yet another mystery.}

{\em{AHA - American Heart Association; NSF - National Science Foundation; VITAL - Ventilator Intervention Technology Accessible Locally.}}

\vskip 15pt

{\hfill
\begin{tabular}{c|c}
  \toprule[1pt] 
  \textbf{Result of 10 trials} & $n$\\
  \midrule[0.5pt]
  supported $\times$ & 0 \\
  partially supported & 1 \\
  not supported \checkmark & 9 \\
  \midrule[0.5pt]
  \multicolumn{2}{c}{\textbf{Truth: Claim is false}} \\
  \bottomrule[1pt]
\end{tabular}
\hfill}
\vskip 5pt
Trials were run using identical prompts on pristine instances of the GPT3.5 chatbot.  Classification of responses into supported {\em etc.} was done manually.  In this small test effectiveness is over 90\%.

\clearpage

\subsection{Simulated persona Feynman's python code}

This code was generated by ChatGPT4 in response to a scenario prompt.  It has been slightly edited for line wrapping and to replace single quotes with double quotes to be more robust for cut and paste.

\small\begin{verbatim}
import numpy as np
import matplotlib.pyplot as plt
def sinc(x):
    return np.sinc(x / np.pi)
def probability_distribution(x, omega, t1, t2, T1, T2, A1, A2):
    term1 = A1**2 * sinc(omega * T1 / 2)**2
    term2 = A2**2 * sinc(omega * T2 / 2)**2
    term3 = 2 * A1 * A2 * sinc(omega * T1 / 2) * \
        sinc(omega * T2 / 2) * np.cos(2 * k * x - omega * (t1 - t2))
    return term1 + term2 + term3
# Parameters
A1 = A2 = 1
T1 = T2 = 1
k = 2 * np.pi
x=0
time_delays = np.linspace(-10, 10, 500)
frequencies = np.linspace(-10, 10, 500)
# Create a meshgrid for time delays and frequencies
time_delay_grid, frequency_grid = np.meshgrid(time_delays, frequencies)
# Compute the probability distribution for each combination
# of time delay and frequency
probability_grid = probability_distribution(x, frequency_grid, 0,
    time_delay_grid, T1, T2, A1, A2)
# Plot the interference pattern over the time-frequency domain
plt.pcolormesh(time_delays, frequencies, probability_grid, cmap="viridis")
plt.xlabel("Time delay (t1 - t2)")
plt.ylabel("Frequency (omega’)")
plt.title("Interference pattern in time-frequency domain")
plt.colorbar(label="Probability distribution")
plt.show()
\end{verbatim}

\begin{bibdiv}
\begin{biblist}

\bib{ma2021scope}{article}{
  title={The Scope of ChatGPT in Software Engineering: A Thorough 	
  Investigation},
  author={Wei Ma{,} Shangqing Liu{,} Wenhan Wang{,} Qiang Hu{,} Ye Liu{,}
  Cen Zhang{,} Liming Nie{,} and Yang Liu},
  journal={arXiv preprint arXiv:2305.12138},
  year={2023},
  url={http://arxiv.org/abs/2305.12138}
}

\bib{akin2023awesome}{misc}{
    author = {Fatih Kadir Akin and 72 Contributors},
    title = {Awesome chatgpt prompts},
    year = {2023},
    note={\url{https://github.com/f/awesome-chatgpt-prompts}}
    }

\bib{xu2023expertprompting}{article}{
  title={ExpertPrompting: Instructing Large Language Models to be 
  Distinguished Experts},
  author={Benfeng Xu{,} An Yang{,} Junyang Lin{,} Quan Wang{,}
  Chang Zhou{,} Yongdong Zhang{,} and Zhendong Mao},
  journal={arXiv preprint arXiv:2305.14688},
  year={2023},
  url={https://arxiv.org/abs/2305.14688}
  }
  
\bib{salewski2023incontext}{article}{
  title={In-Context Impersonation Reveals Large Language
  Models’ Strengths and Biases},
  author={Leonard Salewsi{,} Stephen Alanizi{,} Isabel Rio-Torto{,}
  Eric Schulz{,} and Zeynep Akata},
  journal={arXiv preprint arXiv:2305.14930},
  year={2023},
  url={https://arxiv.orb/abs/2305.14930}
}
  
\bib{du2023improving}{article}{
  title={Improving Factuality and Reasoning in Language Models through 
  Multiagent Debate},
  author={Yilun Du{,} Shuang Li{,} Antonio Torralba{,}
  Joshua B Tenenbaum{,} and Igor Mordatch},
  journal={arXiv preprint arXiv:2305.14325},
  year={2023},
  url={https://composable-models.github.io/llm_debate/}
}

\bib{openaichatgpt4}{article}{
  title={GPT-4 Technical Report},
  author={OpenAI},
  year={2023},
  journal={arXiv preprint arXiv:2303.08774},
  url={https://arxiv.org/pdf/2303.08774.pdf}
}

\bib{park2023generative}{article}{
  title={Generative Agents: Interactive Simulacra of Human Behavior},
  author={Joon Sung Park{,} C. O'Brien{,} Carrie J. Cia{,}
  Meredith Ringel Morris{,} Percy Liang{,} and Michael S. Bernstein},
  journal={arXiv preprint arXiv:2304.03442},
  year={2023}
}

\bib{sawicki2023bits}{article}{
  title={Bits of Grass: Does GPT already know how to write like Whitman?},
  author={Piotr Sawicki{,} Marek Grzes{,} Fabricio Goes{,} Dan Brown{,}
  Max Peeperkorn{,} and Aisha Khatun},
  journal={arXiv preprint arXiv:2305.11064},
  year={2023}
}

\bib{pan2022factchecking}{article}{
  title={Fact-Checking Complex Claims with Program-Guided Reasoning},
  author={Liangming Pan{,} Xiaobao Wu{,} Xinyuan Lu{,} Anh Tuan Luu{,}
  William Yang Wang{,} Min-Yen Kan{,} and Preslav Nakov},
  journal={arXiv preprint arXiv:2305.12744},
  year={2023},
  url={https://arxiv.org/abs/2305.12744}
}

\bib{openaichatapi}{misc}{
  title={API Reference},
  author={OpenAI},
  year={Retrieved 20230526},
  note={\url{https://platform.openai.com/docs/api-reference}}
}

\bib{doyle1930complete}{book}{
  title={The complete Sherlock Holmes},
  author={Doyle, Arthur Conan},
  year={1930},
  publisher={Doubleday Books}
}

\bib{vitalventilatorvideo}{misc}{
  title={NASA Builds Ventilator Prototype for Coronavirus Patients},
  author={NASA Jet Propulsion Laboratory},
  note={\url{https://www.youtube.com/watch?v=NB7SdwkBqHU&t=9s}}
}

\bib{tirole2022double}{article}{
  title={Double-slit time diffraction at optical frequencies},
  author={Romain Tirole{,} Stefano Vezzoli{,} Emanuele Galiffi{,}
  Ian Robertson{,} Dries Maurice{,} Benjamin Tilmann{,} Stefan A Maier{,}
   John B Pendry{,} and Riccardo Sapienza},
  journal={arXiv preprint arXiv:2206.04362},
  url={\url{https://arxiv.org/abs/2206.04362}},
  year={2022}
  note={Figure used with license {https://creativecommons.org/licenses/by/4.0/}, published also in Nature Physics, 2023, Nature Publishing Group UK London}
  
}

\bib{openaiplayground}{misc}{
  title={OpenAI Playground},
  author={{Open}{AI}},
  note={\url{https://platform.openai.com/playground}}
}

\bib{ChatGPT4webpage}{misc}{
  title={Chat GPT4},
  author={OpenAI},
  note={\url{https://chat.openai.com/}}
}

\bib{IAS}{article}{
  title={Mission and History},
  author={Institute for Advanced Study},
  note={\url{https://www.ias.edu/about/mission-history}},
  publisher={Doubleday Books}
}

\bib{wolframalpha}{misc}{
  title={WolframAlpha},
  author={WolframAlpha},
  note={\url{https://www.wolframalpha.com/}}
}

\bib{dalle2}{misc}{
  title={DALL-E},
  author={OpenAI},
  note={\url{https://labs.openai.com}}
}

\end{biblist}
\end{bibdiv}
\end{document}